\documentclass[10pt,conference]{IEEEtran}

\usepackage{amssymb,amsfonts,amsmath,amstext,mathrsfs}
\usepackage{latexsym}
\usepackage{epsfig,psfrag,color,graphics,epsf}
\usepackage{enumerate}

\newtheorem{corollary}{\textbf{Corollary}}
\newtheorem{lemma}{\textbf{Lemma}}
\newtheorem{theorem}{\textbf{Theorem}}

\newtheorem{remark}{\textbf{Remark}}

\newcommand{\pp}{\hspace{1cm}}
\newcommand{\spp}{\hspace{5mm}}

\newcommand{\mE}{\mathrm{E}}

\newcommand{\cX}{\mathcal{X}}
\newcommand{\tcX}{\tilde{\mathcal{X}}}
\newcommand{\cY}{\mathcal{Y}}
\newcommand{\tcY}{\tilde{\mathcal{Y}}}
\newcommand{\cZ}{\mathcal{Z}}
\newcommand{\tcZ}{\tilde{\mathcal{Z}}}

\newcommand{\cW}{\mathcal{W}}

\newcommand{\cCN}{\mathcal{CN}}

\newcommand{\tx}{\tilde{x}}
\newcommand{\ty}{\tilde{y}}
\newcommand{\tz}{\tilde{z}}

\newcommand{\tX}{\tilde{X}}
\newcommand{\tY}{\tilde{Y}}

\newcommand{\tZ}{\tilde{Z}}

\newcommand{\uh}{\underline{h}}

\DeclareMathAlphabet{\matheuf}{U}{euf}{m}{n}

\IEEEoverridecommandlockouts

\begin{document}

\title{Secure Communication over Fading Channels
%
%
\thanks{This research was supported by the National Science Foundation
under Grant ANI-03-38807.} }
\author{\authorblockN{Yingbin Liang and H. Vincent Poor}
\authorblockA{Department of Electrical Engineering \\
Princeton University\\
Princeton, NJ 08544, USA \\
Email: \{yingbinl,poor\}@princeton.edu} }

\maketitle

\thispagestyle{plain}

\begin{abstract}
The fading wire-tap channel is investigated, where the
source-to-destination channel and the source-to-wire-tapper
channel are corrupted by multiplicative fading gain coefficients
in addition to additive Gaussian noise terms. The channel state
information is assumed to be known at both the transmitter and the
receiver. The parallel wire-tap channel with independent
subchannels is first studied, which serves as an
information-theoretic model for the fading wire-tap channel. Each
subchannel is assumed to be a general broadcast channel and is not
necessarily degraded. The secrecy capacity of the parallel
wire-tap channel is established, which is the maximum rate at
which the destination node can decode the source information with
small probability of error and the wire-tapper does not obtain any
information. This result is then specialized to give the secrecy
capacity of the fading wire-tap channel, which is achieved with
the source node dynamically changing the power allocation
according to the channel state realization. An optimal source
power allocation is obtained to achieve the secrecy capacity. This
power allocation is different from the water-filling allocation
that achieves the capacity of fading channels without the secrecy
constraint.
\end{abstract}

\section{Introduction}

Wireless communication has a broadcast nature, where security
issues are captured by a basic wire-tap channel introduced by
Wyner in \cite{Wyner75}. In this model, a source node wishes to
transmit confidential information to a destination node and wishes
to keep a wire-tapper as ignorant of this information as possible.
The performance measure of interest is the secrecy capacity, which
is the largest reliable communication rate from the source node to
the destination node with the wire-tapper obtaining no
information. For the wire-tap channel where the channel from the
source node to the destination and the wire-tapper is degraded,
the secrecy capacity was given in \cite{Wyner75} for the discrete
memoryless channel and in \cite{Leung78} for the Gaussian channel.
The general wire-tap channel without a degradedness assumption and
with an additional common message for both the destination node
and the wire-tapper was considered in \cite{Csiszar78}, where the
capacity-equivocation region and the secrecy capacity were given.
The wire-tap channel was also considered recently for the fading
and multiple antenna channels in \cite{Parada05,Barros06}. The
secrecy capacity was addressed either for the case with a fixed
fading state or from the outage probability viewpoint.

In this paper, we study the ergodic secrecy capacity of the fading
wire-tap channel, which is the maximum secrecy rate that can be
achieved over multiple fading states. We assume the fading gain
coefficients of the source-to-destination channel and the
source-to-wire-tapper channel are stationary and ergodic over
time. We also assume both the transmitter and the receiver know
the channel state information (CSI). Note that the CSI of the
source-to-wire-tapper channel at the source can be justified as
follows. In wireless networks, a node may be treated as a
``wire-tapper" by a source node because it is not the intended
destination of particular confidential messages. In this case, the
``wire-tapper" is not a hostile node, and may also expect its own
information from the same source node. Hence it is reasonable to
assume that this ``wire-tapper" feeds back the CSI to the source
node.


The fading wire-tap channel can be viewed as a special case of the
parallel wire-tap channel with independent subchannels in that the
channel at each fading state realization corresponds to one
subchannel. Hence we first study a parallel wire-tap channel with
$L$ independent subchannels. Each subchannel is assumed to be a
general broadcast channel and is not necessarily degraded, which
is different from the model studied in \cite{Yama89}. This channel
model also differs from the model studied in \cite{Yama89} in that
the wire-tapper can receive outputs from all subchannels. The
secrecy capacity of the parallel wire-tap channel is established.
This result then specializes to the secrecy capacity of a parallel
wire-tap channel with $K+M$ degraded subchannels, which is
directly related to the fading wire-tap channel. For this model,
we assume each of the $K$ subchannels satisfies the condition that
the output at the wire-tapper is a degraded version of the output
at the destination node, and each of the $M$ subchannels satisfies
the condition that the output at the destination node is a
degraded version of the output at the wire-tapper. We show that to
achieve the secrecy capacity, it is optimal to keep the inputs to
the $M$ subchannels null, i.e., use only the $K$ subchannels, and
choose the inputs to the $K$ subchannels independently. Therefore,
the secrecy capacity reduces to the sum of the secrecy capacities
of the $K$ subchannels.


We further apply our result to obtain the secrecy capacity of the
fading wire-tap channel. The fading wire-tap channel we study
differs from the parallel Gaussian wire-tap channel studied in
\cite{Yama91} in that we assume the source node is subject to an
average power constraint over all fading state realization instead
of each subchannel (channel corresponding to one fading state
realization) being subject to a power constraint as assumed in
\cite{Yama91}.
Since the source node knows the CSI, it needs to optimize the
power allocation among fading states to achieve the secrecy
capacity. We obtain the optimal power allocation scheme, where the
source node uses more power when the source-to-destination channel
experiences a larger fading gain and the source-to-wire-tapper
channel has a smaller fading gain. The secrecy capacity is not
achieved by the water-filling allocation that achieves the
capacity for the fading channel without the secrecy constraint.

In this paper, we use $X_{[1,L]}$ to indicate a group of variables
$(X_1,X_2,\ldots,X_L)$, and use $X_{[1,L]}^n$ to indicate a group
of vectors $(X_1^n,X_2^n,\ldots,X_L^n)$, where $X_l^n$ indicates
the vector $(X_{l1},X_{l2},\ldots,X_{ln})$. Throughout the paper,
the logarithmic function is to the base $2$.

The paper is organized as follows. We first introduce the parallel
wire-tap channel with independent subchannels, and present the
secrecy capacity for this channel. We next present the secrecy
capacity for the fading wire-tap channel. We finally demonstrate
our results by numerical examples.

\section{Parallel Wire-tap Channel}\label{sec:model}

\begin{figure*}
\begin{center}
\includegraphics[width=10.5cm,clip]{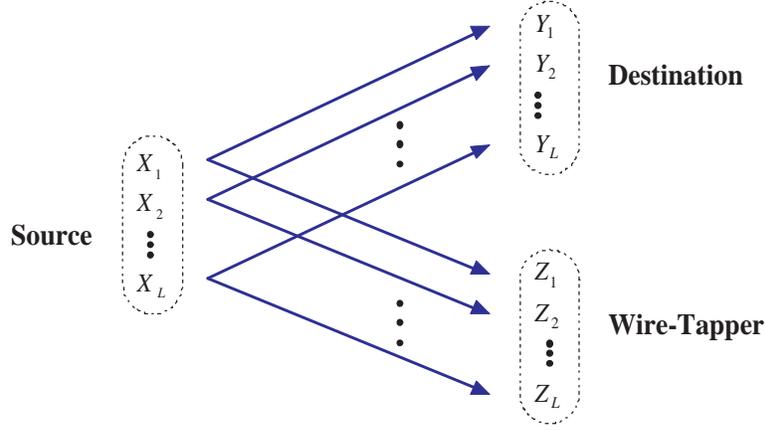}
\caption{Parallel wire-tap channel} \label{fig:wiretap_para}
\end{center}
\end{figure*}

We consider a parallel wire-tap channel with $L$ independent
subchannels (see Fig.~\ref{fig:wiretap_para}), which consists of
$L$ finite input alphabets $\cX_{[1,L]}$, and $2L$ finite output
alphabets $\cY_{[1,L]}$ and $\cZ_{[1,L]}$. The transition
probability distribution is given by
\begin{equation}
p(y_{[1,L]},z_{[1,L]}|x_{[1,L]}) =\prod_{l=1}^L p_l(y_l,z_l| x_l)
\end{equation}
where $x_l \in \cX_l$, $y_l \in \cY_l$, and $z_l \in \cZ_l$ for
$l=1,\ldots,L$.

Note that each of the $L$ subchannels is assumed to be a general
broadcast channel and is not necessarily degraded as assumed in
\cite{Wyner75}. Hence the model we study is more general than the
parallel channel model studied in \cite{Yama89} which assumes each
subchannel is less noisy \cite{Korner77}.

A $\left( 2^{nR}, n \right)$ code consists of the following:
\begin{list}{$\bullet$}{\topsep=0ex \leftmargin=0.4cm
\rightmargin=0.2cm \itemsep=0mm}

\item One message set: $\cW=\{1,2,\ldots,2^{nR}\}$ with the
message $W$ uniformly distributed over $\cW$;

\item One (stochastic) encoder at the source node that maps each
message $w \in \cW $ to a codeword $x_{[1,L]}^n $;

\item One decoder at the destination node that maps a received
sequence $y_{[1,L]}^n $ to a message $w \in \cW$.
\end{list}

The secrecy level of the message $W$ at the wire-tapper is
measured by the {\em equivocation rate} defined as follows:
\begin{equation}
\frac{1}{n} H\left(W \Big|Z_{[1,L]}^n\right).
\end{equation}
The higher the equivocation rate, the less information the
wire-tapper obtains.

A rate-equivocation pair $(R,R_e)$ is {\em achievable} if there
exists a sequence of $\left(2^{nR}, n \right)$ codes with the
destination decoding error probability $P_e^{(n)}\rightarrow 0$ as
$n$ goes to infinity and with the equivocation rate $R_e$
satisfying
\begin{equation}
R_e \leq \lim_{n\rightarrow \infty} \frac{1}{n} H\left(W
\Big|Z_{[1,L]}^n \right).
\end{equation}

We focus on the case where perfect secrecy is achieved, i.e., the
wire-tapper does not obtain any information about the message $W$.
This happens if $R_e=R$. The {\em secrecy capacity} $C_s$ is the
maximum $R$ such that $(R,R_e=R)$ is achievable, i.e.,
\begin{equation}
C_s=\max_{\text{Achievable } (R,R_e=R)} R.
\end{equation}

We obtain the following secrecy capacity result for the parallel
wire-tap channel.
\begin{theorem}\label{th:paracapa}
The secrecy capacity of the parallel wire-tap channel with $L$
subchannels is
\begin{equation}\label{eq:paracapa}
C_s=\sum_{l=1}^L C_s^l
\end{equation}
where $C_s^l$ is the secrecy capacity of subchannel $l$ and is
given by
\begin{equation}\label{eq:csl}
C_s^l=\max \; I(U_l;Y_l)-I(U_l;Z_l)
\end{equation}
where the maximum in the preceding equation is over the
distributions $p(u_l,x_l)p(y_l,z_l|x_l)$, which satisfies the
Markov chain condition $U_l \rightarrow X_l \rightarrow
(Y_l,Z_l)$.
\end{theorem}

The proof of Theorem \ref{th:paracapa} is relegated to Section
\ref{sec:proof}.

In the following, we consider a parallel wire-tap channel, where
each subchannel is either degraded such that the output at the
wire-tapper is a degraded version of the output at the destination
node, or degraded such that the output at the destination node is
a degraded version of the output at the wire-tapper. This channel
specializes to the fading wiretap channel that is considered in
Section \ref{sec:fading}, and is hence of particular interest.

\begin{figure*}
\begin{center}
\includegraphics[width=10.5cm,clip]{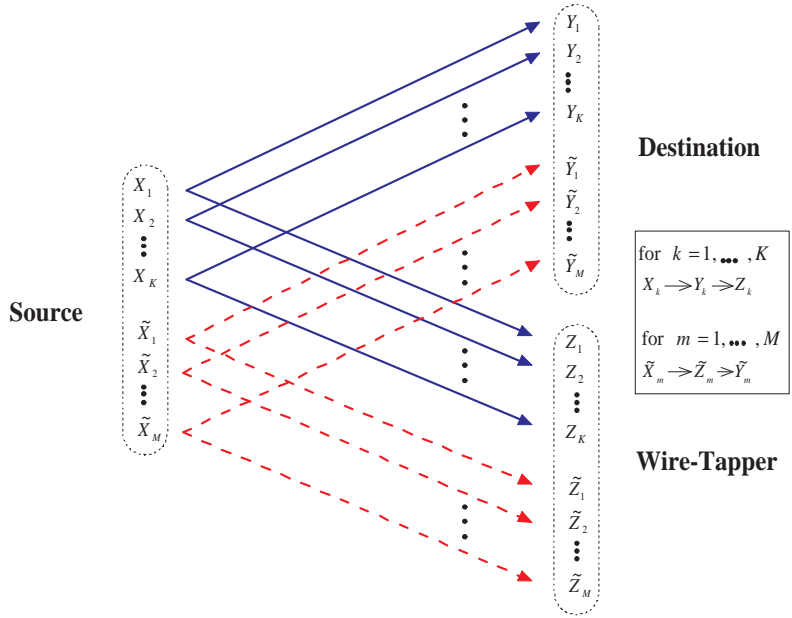}
\caption{Parallel wire-tap channel with $K+M$ degraded
subchannels} \label{fig:wiretap_para_km1}
\end{center}
\end{figure*}

More formally, we define the channel described above to be the
parallel wire-tap channel with $K+M$ degraded subchannels (see
Fig.~\ref{fig:wiretap_para_km1}), which consists of $K+M$ finite
input alphabets $\cX_{[1,K]}$ and $\tcX_{[1,M]}$, $2(K+M)$ finite
output alphabets $\cY_{[1,K]}, \cZ_{[1,K]},\tcY_{[1,M]},$ and
$\tcZ_{[1,M]}$. The transition probability distribution is given
by
\begin{equation}
\begin{split}
&p(y_{[1,K]},\ty_{[1,M]},z_{[1,K]},\tz_{[1,M]}|x_{[1,K]},\tx_{[1,M]})
\\
&\spp =\prod_{k=1}^K p_k(y_k,z_k| x_k) \prod_{m=1}^M
p_m(\ty_m,\tz_m|\tx_m)
\end{split}
\end{equation}
where $x_k \in \cX_k$, $y_k \in \cY_k$, and $z_k \in \cZ_k$ for
$k=1,\ldots,K$, and $\tx_m \in \tcX_m$, $\ty_m \in \tcY_m$, and
$\tz_m \in \tcZ_m$ for $m=1,\ldots,M$. The probability
distributions $p_k(y_k,z_k| x_k)$ and $ p_m(\ty_m,\tz_m|\tx_m)$
satisfy the following degraded conditions:
\begin{equation}\label{eq:dgcond}
\begin{split}
& p_k(y_k,z_k| x_k)= p_k(y_k|x_k)p_k(z_k|y_k) \\
& \hspace{4cm} \text{for } k=1,\ldots,K, \\
& p_m(\ty_m,\tz_m| \tx_m) =p_m(\tz_m|\tx_m)p_m(\ty_m|\tz_m) \\
& \hspace{4cm} \text{for } m=1,\ldots,M;
\end{split}
\end{equation}
i.e., the following Markov chain conditions are satisfied
\begin{equation}\label{eq:dgmark}
\begin{split}
X_k \rightarrow & Y_k \rightarrow Z_k \hspace{1.2cm} \text{for } k=1,\ldots,K, \\
\tX_m \rightarrow & \tZ_m \rightarrow \tY_m \pp \text{for }
m=1,\ldots,M.
\end{split}
\end{equation}

The following secrecy capacity follows easily from Theorem
\ref{th:paracapa}.
\begin{corollary}\label{cor:dgcapa}
The secrecy capacity of the parallel wire-tap channel with $K+M$
degraded subchannels is
\begin{equation}\label{eq:dgcapa}
C_s=\sum_{k=1}^K C_s^k
\end{equation}
where $C_s^k$ is the secrecy capacity of subchannel $k$ and is
given by
\begin{equation}\label{eq:csk}
C_s^k=\max_{p(x_k)} \; I(X_k;Y_k)-I(X_k;Z_k).
\end{equation}
\end{corollary}
\begin{remark}\label{rm:indepinput}
It is optimal to choose the inputs to the $K$ subchannels
independently and set the inputs to the $M$ subchannels to be
null. Hence the $M$ subchannels do not contribute to the secrecy
capacity. This is intuitive because the wire-tapper obtains all
information that the destination node obtains over the $M$
subchannels.
\end{remark}


\section{Proof of Theorem \ref{th:paracapa}}\label{sec:proof}

The achievability follows from \cite[Corollary~2]{Csiszar78} by
setting $U=(U_1,\ldots,U_L)$, $X=(X_1,\ldots,X_L)$,
$Y=(Y_1,\ldots,Y_L)$, and $Z=(Z_1,\ldots,Z_L)$, and choosing the
components of $U$ and $X$ to be independent.

To show the converse, we consider a code with length $n$ and
average error probability $P_e$. The probability distribution on
$W \times \cX_{[1,L]}^n \times \cY_{[1,L]}^n \times \cZ_{[1,L]}^n$
is given by
\begin{equation}
\begin{split}
& p(w,x_{[1,L]}^n,y_{[1,L]}^n,z_{[1,L]}^n) \\
& \spp=p(w)p(x_{[1,L]}^n|w)\prod_{i=1}^n \;\prod_{l=1}^L \;
p_l(y_{li},z_{li}|x_{li})
\end{split}
\end{equation}

By Fano's inequality \cite[Sec.~2.11]{Cover91}, we have
\begin{equation}
H(W|Y_{[1,L]}^n) \leq nR P_e+1 :=n\delta
\end{equation}
where $\delta \rightarrow 0$ if $P_e \rightarrow 0$.

We now bound the equivocation rate $R_e$:
\begin{equation}\label{eq:re}
\begin{split}
& nR_e \\
& \leq H\big(W|Z_{[1,L]}^n\big) \\
& = H\big(W|Z_{[1,L]}^n\big)-H(W)+H(W) \\
& \spp -H\big(W|Y_{[1,L]}^n\big)+H\big(W|Y_{[1,L]}^n\big)\\
& \overset{(a)}{\leq} I(W;Y_{[1,L]}^n)-I(W;Z_{[1,L]}^n)+n\delta \\
& = \sum_{l=1}^L \Big[I(W;Y_l^n|Y_{[1,l-1]}^n)-I(W;Z_l^n|Z_{[l+1,L]}^n)\Big]+n\delta \\
& = \sum_{l=1}^L\sum_{i=1}^n \Big[I(W;Y_{li}|Y_{[1,l-1]}^nY_l^{i-1})\\
& \pp \spp -I(W;Z_{li}|Z_{l[i+1]}^nZ_{[l+1,L]}^n)\Big]+n\delta \\
& \overset{(b)}{=} \sum_{l=1}^L\sum_{i=1}^n \Big[I(WZ_{l[i+1]}^nZ_{[l+1,L]}^n;Y_{li}|Y_{[1,l-1]}^nY_l^{i-1})\\
& \pp \spp -I(Z_{l[i+1]}^nZ_{[l+1,L]}^n;Y_{li}|WY_{[1,l-1]}^nY_l^{i-1})\\
& \pp \spp -I(WY_{[1,l-1]}^nY_l^{i-1};Z_{li}|Z_{l[i+1]}^nZ_{[l+1,L]}^n) \\
& \pp \spp +I(Y_{[1,l-1]}^nY_l^{i-1};Z_{li}|WZ_{l[i+1]}^nZ_{[l+1,L]}^n)\Big]+n\delta \\
& \overset{(c)}{=} \sum_{l=1}^L\sum_{i=1}^n \Big[I(WZ_{l[i+1]}^nZ_{[l+1,L]}^n;Y_{li}|Y_{[1,l-1]}^nY_l^{i-1})\\
& \pp \spp -I(WY_{[1,l-1]}^nY_l^{i-1};Z_{li}|Z_{l[i+1]}^nZ_{[l+1,L]}^n)\Big]+n\delta \\
& = \sum_{l=1}^L\sum_{i=1}^n \Big[I(Z_{l[i+1]}^nZ_{[l+1,L]}^n;Y_{li}|Y_{[1,l-1]}^nY_l^{i-1})\\
& \pp \spp +I(W;Y_{li}|Y_{[1,l-1]}^nY_l^{i-1}Z_{l[i+1]}^nZ_{[l+1,L]}^n)\\
& \pp \spp -I(Y_{[1,l-1]}^nY_l^{i-1};Z_{li}|Z_{l[i+1]}^nZ_{[l+1,L]}^n) \\
& \pp \spp -I(W;Z_{li}|Y_{[1,l-1]}^nY_l^{i-1}Z_{l[i+1]}^nZ_{[l+1,L]}^n)\Big]+n\delta \\
& \overset{(d)}{=} \sum_{l=1}^L\sum_{i=1}^n \big[I(W;Y_{li}|Y_{[1,l-1]}^nY_l^{i-1}Z_{l[i+1]}^nZ_{[l+1,L]}^n)\\
& \pp \spp -I(W;Z_{li}|Y_{[1,l-1]}^nY_l^{i-1}Z_{l[i+1]}^nZ_{[l+1,L]}^n)\Big]+n\delta \\
& \overset{(e)}{=} \sum_{l=1}^L\sum_{i=1}^n\Big[ I(U_{li};Y_{li}|Q_{li})-I(U_{li};Z_{li}|Q_{li}) \Big]+n\delta\\
\end{split}
\end{equation}
where $(a)$ follows from Fano's inequality, $(b)$ follows from the
chain rule, $(c)$ and $(d)$ follow from Lemma~7 in
\cite{Csiszar78}, and $(e)$ follows from the following definition:
\begin{equation}
Q_{li}:=(Y_{[1,l-1]}^nY_l^{i-1}Z_{l[i+1]}^nZ_{[l+1,L]}^n), \spp
U_{li}=(WQ_{li}).
\end{equation}
We note that $(Q_{li},U_{li},X_{li},Y_{li},Z_{li})$ satisfy the
following Markov chain condition:
\begin{equation}
Q_{li}\rightarrow U_{li} \rightarrow X_{li} \rightarrow
(Y_{li},Z_{li}).
\end{equation}

We introduce a random variable $G$ that is independent of all
other random variables, and is uniformly distributed over $\{ 1,2,
\ldots, n\}$. Define $Q_l=(G,Q_{lG})$, $U_l=(G,U_{lG})$,
$X_l=X_{lG}$, $Y_l=Y_{lG}$, and $Z_l=Z_{lG}$. Note that
$(Q_l,U_l,X_l,Y_l,Z_l)$ satisfy the following Markov chain
condition:
\begin{equation}
Q_l\rightarrow U_l \rightarrow X_l \rightarrow (Y_l,Z_l).
\end{equation}

Using the above definitions, \eqref{eq:re} becomes
\begin{equation}\label{eq:re1}
R_e \leq \sum_{l=1}^L \Big[I(U_l;Y_l|Q_l)-I(U_l;Z_l|Q_l)\Big]+
\delta
\end{equation}

Therefore, an upper bound on $R_e$ is
\begin{equation}\label{eq:re2}
R_e \leq \max \; \sum_{l=1}^L
\Big[I(U_l;Y_l|Q_l)-I(U_l;Z_l|Q_l)\Big]+ \delta
\end{equation}
where the maximum is over the probability distributions
$p(q_{[1,L]},u_{[1,L]},x_{[1,L]},y_{[1,L]},z_{[1,L]})$. Finally,
we note that each term in the summation in \eqref{eq:re2} depends
only on the distribution $p(q_l,u_l,x_l,y_l,z_l)$. Hence there is
no loss of optimality to consider only the distributions with the
form $\prod_{l=1}^L p(q_l,u_l,x_l)p(y_l,z_l|x_l)$. We also note
that each term in the summation in \eqref{eq:re2} is maximized by
a constant $Q_l$. Hence the following bound does not lose
optimality:
\begin{equation}
R_e \leq \sum_{l=1}^L \max \Big[I(U_l;Y_l)-I(U_l;Z_l)\Big]+ \delta
\end{equation}
where the maximum for the $l$-th term in the summation is over the
distributions $p(u_l,x_l)p(y_l,z_l|x_l)$ for $l=1,\ldots,L$. This
concludes the converse proof.
\section{Fading Wire-tap Channel}\label{sec:fading}

\begin{figure}[tbhp]
\begin{center}
\includegraphics[width=8cm]{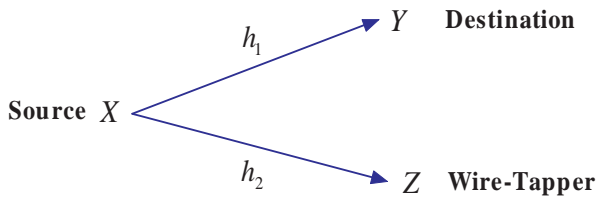}
\caption{Fading Wire-tap Channel} \label{fig:fdwire}
\end{center}
\end{figure}

We study the fading wire-tap channel (see Fig.~\ref{fig:fdwire}),
where the source-to-destination channel and the
source-to-wire-tapper channel are corrupted by multiplicative
fading processes in addition to additive white Gaussian processes.
The channel input-output relationship is given by
\begin{equation}\label{eq:fdmodel}
\begin{split}
Y_i & = h_{1i} X_i+W_i, \\
Z_i & = h_{2i} X_i+V_i,
\end{split}
\end{equation}
where $i$ is the time index, and $X_i$ is the channel input at the
time instant $i$, and $Y_i$ and $Z_i$ are channel outputs at the
time instant $i$, respectively. The channel gain coefficients
$h_{1i}$ and $h_{2i}$ are zero-mean proper complex random
variables. We define $\uh_i:=(h_{1i},h_{2i})$, and assume
$\{\uh_i\}$ is a stationary and ergodic vector random process. The
noise processes $\{W_i\}$ and $\{V_i\}$ are zero-mean independent
identically distributed (i.i.d.) proper complex Gaussian with
$W_i$ and $V_i$ having variances $\mu^2$ and $\nu^2$,
respectively. The input sequence $\{X_i \}$ is subject to the
average power constraint $P$, i.e., $\frac{1}{n}\sum_{i=1}^n
\mE\big[ X^2_i \big]\leq P$. We assume the channel state
information (realization of $\uh_i$) is known at both the
transmitter and the receiver instantaneously. As we mentioned in
the introduction, the source node gets the CSI of the channel to
the wire-tapper when the wire-tapper is not an actual hostile node
and is only not the intended destination node for a particular
confidential message.

We first introduce the following lemma that follows from
\cite[lemma~1]{Liang06isit}. This lemma is useful to obtain the
secrecy capacity of the fading wire-tap channel.
\begin{lemma}\label{lemma:marg}
The secrecy capacity of the wire-tap channel depends only on the
marginal transition distributions $p(y|x)$ of the
source-to-destination channel and $p(z|x)$ of the
source-to-wire-tapper channel.
\end{lemma}

The following generalization of the result in \cite{Leung78}
follows directly from Lemma \ref{lemma:marg}.
\begin{corollary}
The secrecy capacity of the Gaussian wire-tap channel given in
\cite[Theorem~1]{Leung78} holds for the case with general
correlation between the noise variables at the destination node
and the wire-tapper.
\end{corollary}

Based on Lemma \ref{lemma:marg} and Corollary \ref{cor:dgcapa}, we
obtain the secrecy capacity of the fading wire-tap channel.
\begin{theorem}\label{th:fdcapa}
The secrecy capacity of the fading wire-tap channel is
\begin{equation}\label{eq:fdcapa}
\begin{split}
C_s=\max_{\mE_A [P(\uh)] \leq P } \; \mE_A \Bigg[
& \log\left(1+\frac{P(\uh)|h_1|^2}{\mu^2}\right) \\
& -\log\left(1+\frac{P(\uh)|h_2|^2}{\nu^2}\right)\Bigg].
\end{split}
\end{equation}
where $A:= \Big\{\uh: \frac{|h_1|^2}{\mu^2} >
\frac{|h_2|^2}{\nu^2} \Big\}$.
The random vector $\uh=(h_1,h_2)$ has the same distribution as the
marginal distribution of the process $\{\uh_i\}$ at one time
instant.

The optimal power allocation that achieves the secrecy capacity in
\eqref{eq:fdcapa} is given by
\begin{equation}\label{eq:optpower}
P^*(\uh)=
\begin{cases}
\frac{1}{\lambda\ln 2}-\frac{\mu^2}{|h_1|^2}, \spp \text{if } \;
|h_2|^2=0, \;\; \lambda <\frac{1}{\ln
2}\frac{|h_1|^2}{\mu^2} \\ \\
\frac{1}{2}\sqrt{\left(\frac{\nu^2}{|h_2|^2}-\frac{\mu^2}{|h_1|^2}\right)
\left(\frac{4}{\lambda\ln
2}-\frac{\mu^2}{|h_1|^2}+\frac{\nu^2}{|h_2|^2}\right)} \\
-\frac{1}{2}\left(\frac{\mu^2}{|h_1|^2}+\frac{\nu^2}{|h_2|^2}\right),
\\ \\
\hspace{2.6cm} \text{if }\; |h_2|^2 > 0,\;\; \frac{|h_1|^2}{\mu^2}
> \frac{|h_2|^2}{\nu^2}, \\
\hspace{3.1cm} \lambda <\frac{1}{\ln
2}\left(\frac{|h_1|^2}{\mu^2}-\frac{|h_2|^2}{\nu^2}\right) \\ \\
0, \hspace{2.2cm} \text{otherwise}
\end{cases}
\end{equation}
where $\lambda$ is chosen to satisfy the power constraint $\mE_A[
P(\uh)]=P$.
\end{theorem}
\begin{remark}
The optimal power allocation \eqref{eq:optpower} to achieve the
secrecy capacity is not water-filling. This is in contrast to the
fading channel without the secrecy constraint where water-filling
allocation is optimal to achieve the capacity \cite{Gold97}.
\end{remark}
\begin{remark}
The secrecy capacity in Theorem \ref{th:fdcapa} is established for
general fading processes $\{\uh_i\}$ where only ergodic and
stationary conditions are assumed. The fading process $\{\uh_i\}$
can be correlated across time, and is not necessarily Gaussian.
The two component processes $\{h_{1i}\}$ and $\{h_{2i}\}$ can be
correlated as well.
\end{remark}
\begin{remark}
The secrecy capacity in Theorem \ref{th:fdcapa} is established for
the case with general correlation between the noise variables
$W_i$ and $V_i$.
\end{remark}

\begin{figure}
\begin{center}
\begin{tabular}{c}
\includegraphics[width=7.5cm]{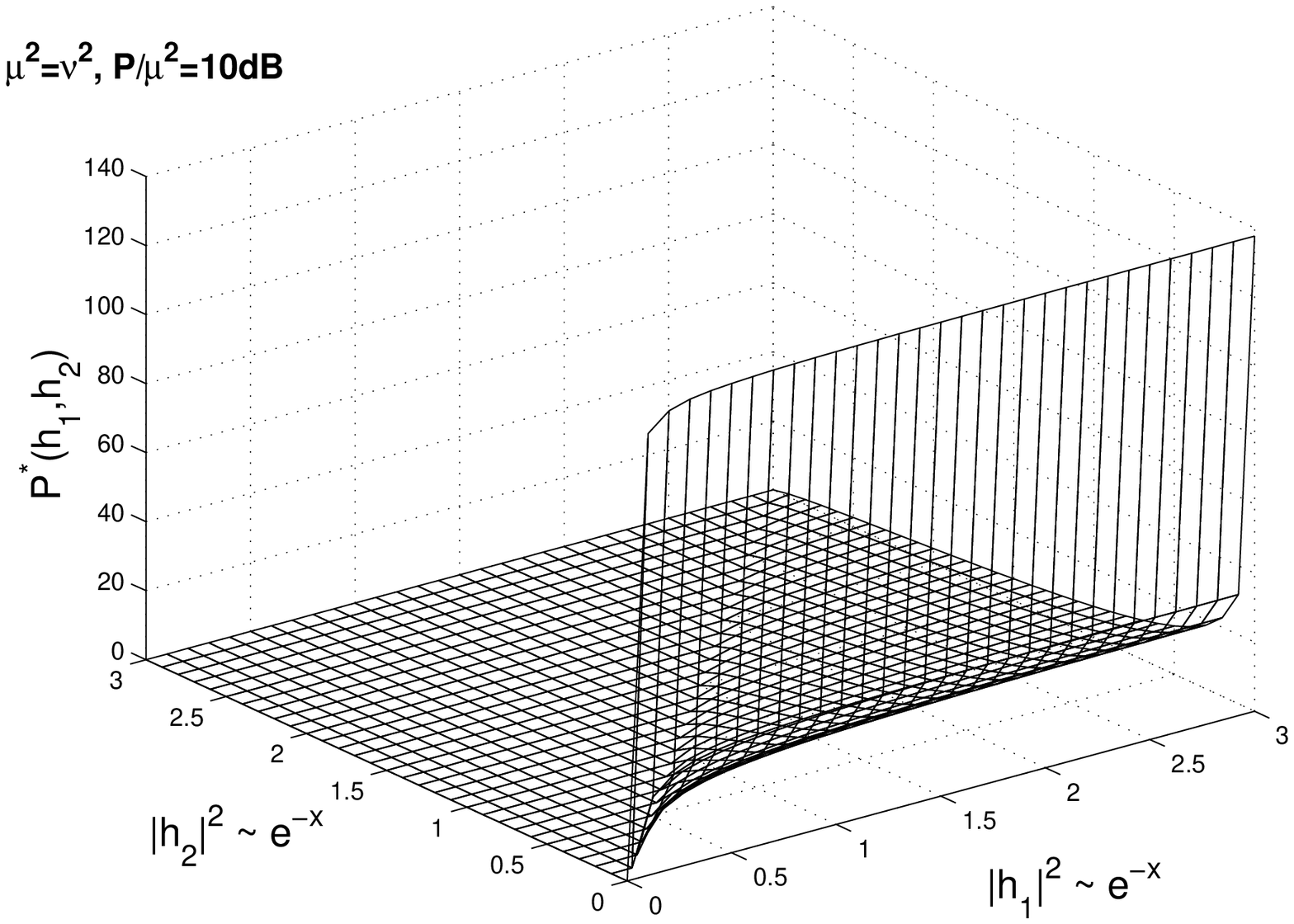}
\\ \\
(a): $P^*(\uh)$ as a function of $(h_1,h_2)$
\\ \\
\includegraphics[width=7.5cm]{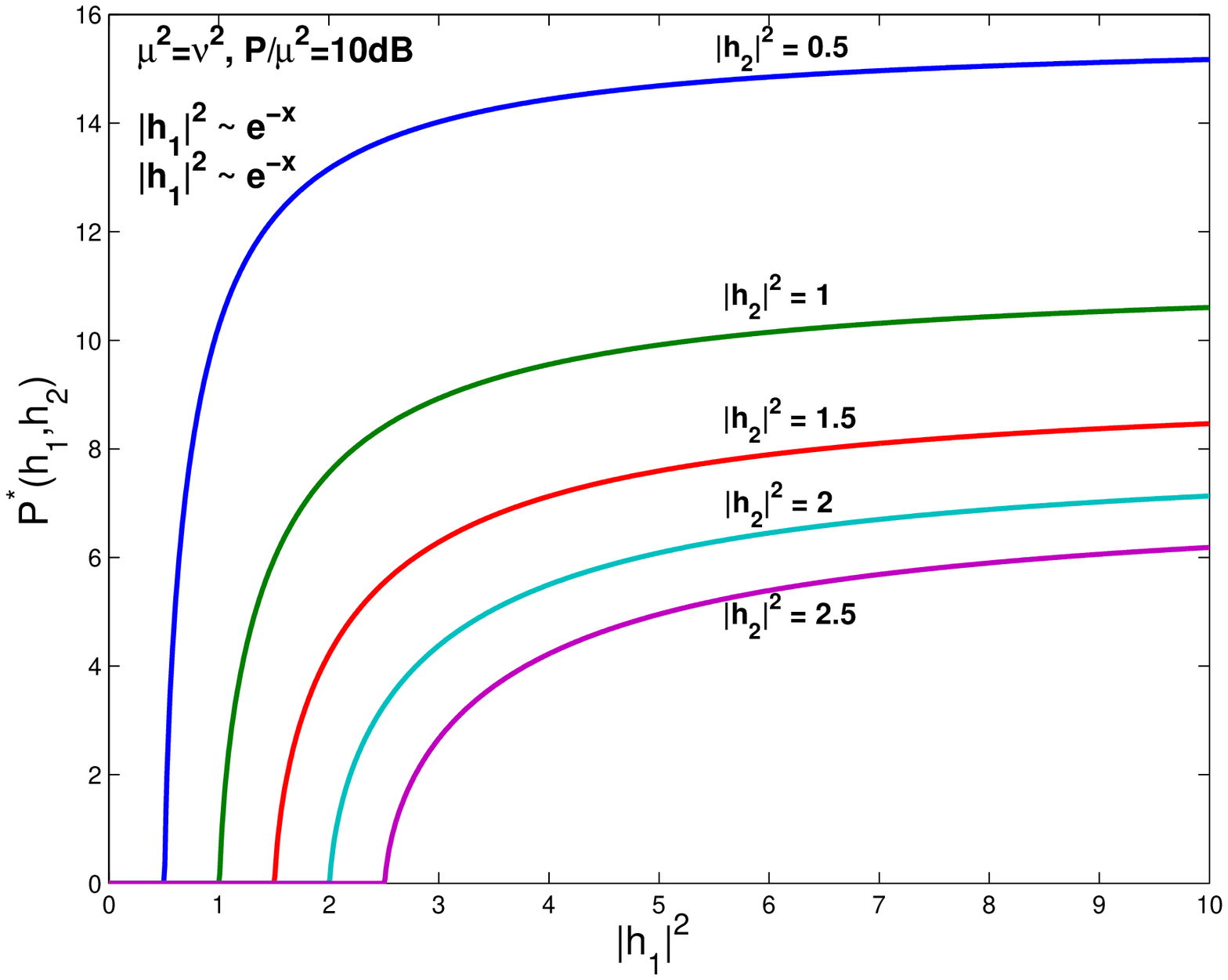}
\\ \\
(b): $P^*(\uh)$ as a function of $|h_1|^2$
\\ \\
\includegraphics[width=7.5cm]{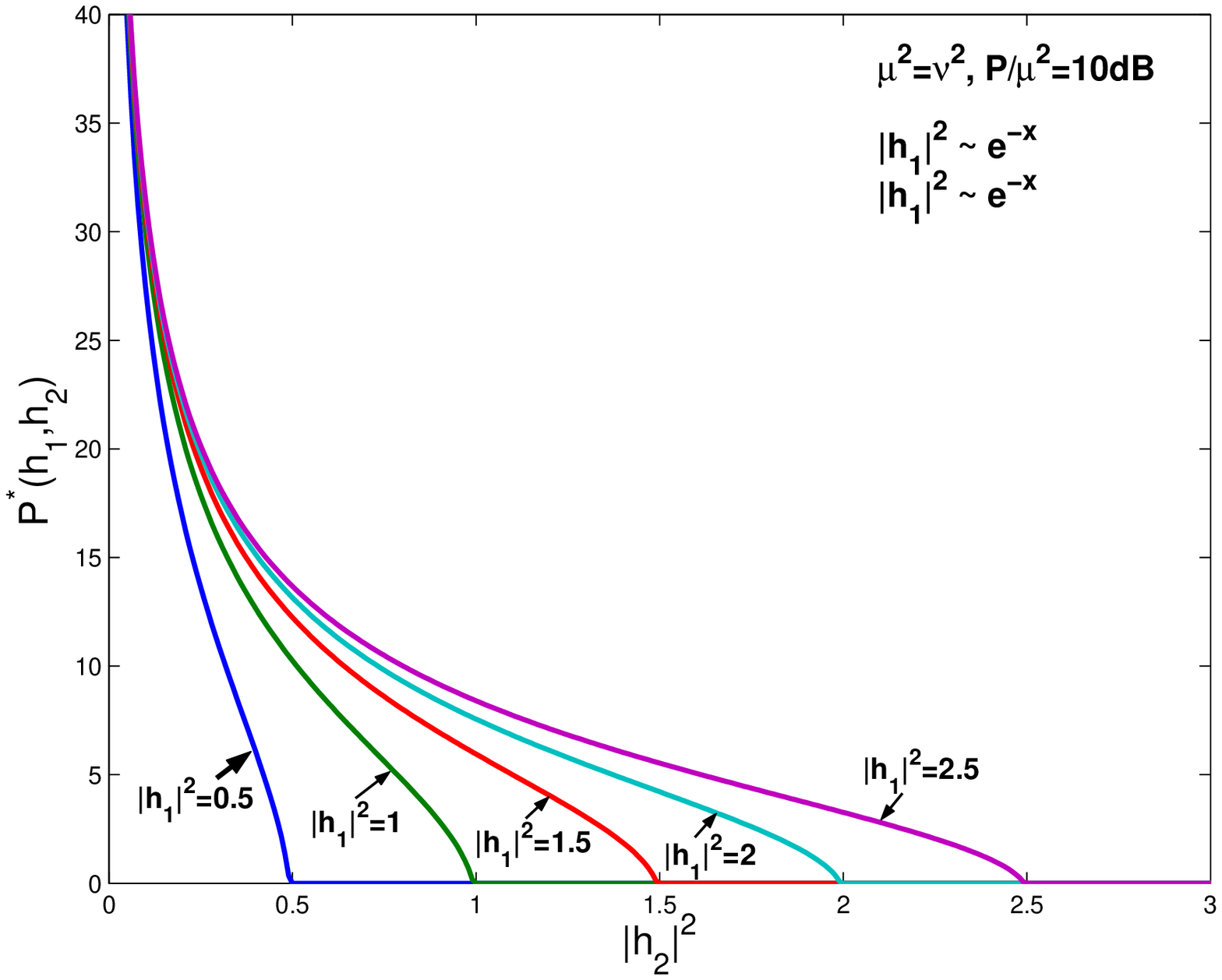}
\\ \\
(c): $P^*(\uh)$ as a function of $|h_2|^2$
\end{tabular}
\caption{Optimal power allocation $P^*(\uh)$ for a Rayleigh fading
wire-tap channel} \label{fig:powerdist}
\end{center}
\end{figure}

\begin{proof}
The fading wire-tap channel can be viewed as a parallel wire-tap
channel with each subchannel having the following form
\begin{equation}\label{eq:fdmodel2}
\begin{split}
Y & = h_1 X+W, \\
Z & = h_2 X+V,
\end{split}
\end{equation}
where $(h_1,h_2)$ is a fixed channel realization of $\uh$. Note
that the subchannel \eqref{eq:fdmodel2} is not physically
degraded. We now consider the following subchannel:
\begin{eqnarray}
&Y = h_1 X+W,\spp Z = \frac{h_2h_1^*}{|h_1|^2}(h_1 X+W)+ V', \nonumber\\
& \hspace{5.5cm} \text{ if }\; \uh \in A \label{eq:fdmodel3}  \\
&Y = \frac{h_1h_2^*}{|h_2|^2}(h_2 X+V)+ W', \spp Z = h_2 X+V, \nonumber \\
& \hspace{5.6cm} \text{ if }\; \uh \in A^c \label{eq:fdmodel4}
\end{eqnarray}
where $V'$ and $W'$ are zero mean proper complex Gaussian random
variables with variances $\nu^2-\frac{|h_2|^2}{|h_1|^2}\mu^2$ and
$\mu^2-\frac{|h_1|^2}{|h_2|^2}\nu^2$, respectively. The subchannel
\eqref{eq:fdmodel3}/\eqref{eq:fdmodel4} is physically degraded,
and has the same marginal distribution $p(y|x)$ and $p(z|x)$ as
the subchannel \eqref{eq:fdmodel2}. Hence by Lemma
\ref{lemma:marg}, the parallel wire-tap channel with subchannels
having the form \eqref{eq:fdmodel2} and with subchannels having
the form \eqref{eq:fdmodel3}/\eqref{eq:fdmodel4} have the same
secrecy capacity. We can now apply Corollary \ref{cor:dgcapa} to
the parallel wire-tap channel with subchannels having the form
\eqref{eq:fdmodel3}/\eqref{eq:fdmodel4}. Note that the subchannel
\eqref{eq:fdmodel3} with $\uh\in A$ is degraded in the same
fashion as the $K$ subchannels in \eqref{eq:dgcond}, and the
subchannel \eqref{eq:fdmodel4} with $\uh \in A^c$ is degraded in
the same fashion as the $M$ subchannels in \eqref{eq:dgcond}. From
Corollary \ref{cor:dgcapa}, it is clear that the subchannels with
$\uh \in A^c$ do not contribute to the secrecy capacity. The
achievability of \eqref{eq:fdcapa} now follows from
\eqref{eq:dgcapa} and \eqref{eq:csk} by setting the input
distribution $X \sim \cCN(0,P(\uh))$ for $\uh \in A$. Note that
the summation $\sum_{k=1}^{K}$ in \eqref{eq:dgcapa} becomes the
average $\mE_{\uh \in A}$ for the fading channel.

The converse of \eqref{eq:fdcapa} follows from the steps that are
similar to those in \cite{Leung78}.

We are now left to optimize \eqref{eq:fdcapa} over power
allocations satisfying $\mE_A[P(\uh)] \leq P$. One can check that
the following function of $P(\uh)$
\begin{equation}
\mE_A \left[
\log\left(1+\frac{P(\uh)|h_1|^2}{\mu^2}\right)-\log\left(1+\frac{P(\uh)|h_2|^2}{\nu^2}\right)\right]
\end{equation}
is concave. The optimal $P^*(\uh)$ given in \eqref{eq:optpower}
can be derived by the standard Kuhn-Tucker condition (see e.g.,
\cite[p.~314-315]{Luen03}).
\end{proof}
\section{Numerical Results}

We first consider the Rayleigh fading wire-tap channel, where
$h_1$ and $h_2$ are zero mean proper complex Gaussian random
variables with variances 1. Hence $|h_1|^2$ and $|h_2|^2$ are
exponentially distributed with parameter $1$. In
Fig.~\ref{fig:powerdist} (a), we plot the optimal power allocation
$P^*(\uh)$ as a function of $\uh$. It can be seen from the graph
that most of the source power is allocated to the channel states
with small $|h_2|^2$. This behavior is shown more clearly in
Fig.~\ref{fig:powerdist} (b), which plots $P^*(\uh)$ as a function
of $|h_1|^2$ for different values of $|h_2|^2$, and in
Fig.~\ref{fig:powerdist} (c), which plots $P^*(\uh)$ as a function
of $|h_2|^2$ for different values of $|h_1|^2$. The source node
allocates more power to the channel states with larger $|h_1|^2$
to forward more information to the destination node, and allocates
less power for the channel states with larger $|h_2|^2$ to prevent
the wire-tapper to obtain information. It can also be seen from
Fig.~\ref{fig:powerdist} (b) and Fig.~\ref{fig:powerdist} (c) that
the source node transmits only when the source-to-destination
channel is better than the source-to-wire-tapper channel.

\begin{figure}
\begin{center}
\includegraphics[width=7.5cm]{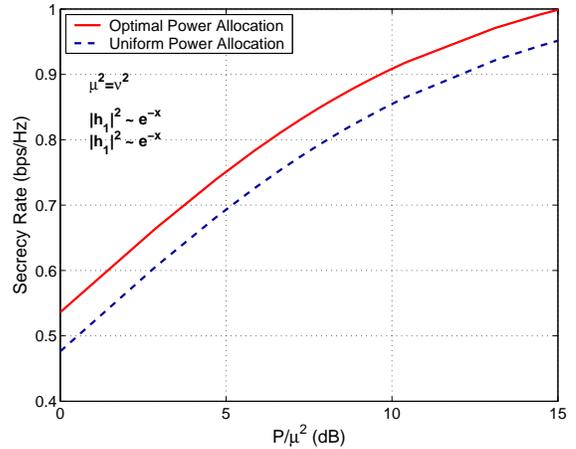}
\caption{Comparison of secrecy capacity by optimal power
allocation with secrecy rate by uniform power allocation for a
Rayleigh fading wire-tap channel} \label{fig:crcs_sig1}
\end{center}
\end{figure}

Fig.~\ref{fig:crcs_sig1} plots the secrecy capacity achieved by
the optimal power allocation, and compares it with the secrecy
rate achieved by a uniform power allocation, i.e., allocating the
same power for all channel states $\uh \in A$. It can be seen that
the uniform power allocation does not provide performance close to
the secrecy capacity for the SNRs of interest. This is in contrast
to the Rayleigh fading channel without the secrecy constraint,
where the uniform power allocation can be close to optimum even
for moderate SNRs. This also demonstrates that the exact channel
state information is important to achieve higher secrecy rate.

\begin{figure}
\begin{center}
\includegraphics[width=7.5cm]{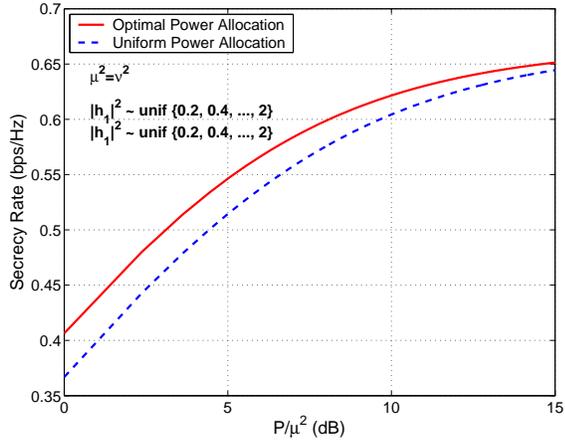}
\caption{Comparison of secrecy capacity by optimal power
allocation with secrecy rate by uniform power allocation for a
uniformly distributed fading wire-tap channel}
\label{fig:crcs_unif}
\end{center}
\end{figure}

We next consider a fading wire-tap channel, where $|h_1|^2$ and
$|h_2|^2$ are uniformly distributed over finite mass points
$\{0.2, 0.4, \ldots, 2\}$. It can be seen from
Fig.~\ref{fig:crcs_unif} that the secrecy rate achieved by the
uniform power allocation approaches the secrecy capacity as SNR
increases. Hence the uniform power allocation can be close to
optimum for certain distributions of the fading gain coefficients.

\section{Conclusions}

We have established the secrecy capacity for the parallel wire-tap
channel with independent subchannels. We have further applied this
result to obtain the secrecy capacity for the fading wire-tap
channel, where the channel state information is assumed to be
known at both the transmitter and the receiver. In particular, we
have derived the optimal power allocation scheme to achieve the
secrecy capacity. Our numerical results demonstrate that the
channel state information at the transmitter is useful to improve
the secrecy capacity.

\bibliographystyle{IEEEtran}

\end{document}